\documentclass{pnastwo}
\usepackage[dvips]{graphicx}
\usepackage{amssymb,amsfonts,amsmath}
\url{www.pnas.org/cgi/doi/10.1073/pnas.0709640104}
\copyrightyear{2008}
\issuedate{Issue Date}
\volume{Volume}
\issuenumber{Issue Number}

\begin{document}

\title{Iron pnictides as a new setting for quantum criticality}

\author{Jianhui Dai\affil{1}{Zhejiang Institute of Modern Physics, 
Zhejiang University, Hangzhou 310027, China},
Qimiao Si\affil{2}{Department of Physics and Astronomy, Rice University,
Houston, Texas 77005, USA},
Jian-Xin Zhu\affil{3}{Theoretical Division, Los Alamos National Laboratory, Los Alamos, New Mexico 87545, USA},
Elihu Abrahams\affil{4}{Center for Materials Theory, Department of Physics and Astronomy, Rutgers University, 
Piscataway, New Jersey 08855, USA}}
\contributor{Submitted to Proceedings of the National Academy of Sciences
of the United States of America}

\maketitle

\begin{article}
\begin{abstract}
Two major themes in the physics of condensed matter are quantum critical
phenomena and unconventional superconductivity. These usually occur 
in the context of competing interactions in systems of strongly-correlated
electrons.  All this interesting physics comes together in the behavior 
of the recently discovered iron pnictide compounds that have generated 
enormous interest because of their moderately high-temperature 
superconductivity. The ubiquity of antiferromagnetic ordering in their
phase diagrams naturally raises the question of the relevance of magnetic
quantum criticality, but the answer remains uncertain both theoretically and experimentally. Here we show that the undoped iron pnictides feature 
a novel type of magnetic 
quantum critical point, which results from a competition between electronic localization and itinerancy. Our theory provides a mechanism to understand the experimentally-observed variation of the ordered moment among the undoped iron pnictides. We suggest P substitution for As in the undoped iron pnictides as a means to access this new example of magnetic quantum criticality in an unmasked fashion. Our findings point to the iron pnictides as a much-needed new setting for quantum criticality, one that offers a new set of control parameters.\end{abstract}


\section{\dropcap{Q}uantum criticality in the pnictides}
\ \ \ \ \ \ \ \ The recent discovery of copper-free high-$T_c$ 
superconductors has triggered intense interest in the homologous 
iron pnictides. The
parent compound of the lanthanum-iron oxyarsenide,
LaOFeAs~\cite{Kamihara:JACS08}, exhibits a 
tetragonal-orthorhombic
structural
transition  and  long-range
antiferromagnetic order~\cite{Cruz:08}.
Electron doping, via fluorine substitution for oxygen,
suppresses both and induces superconductivity. Other families of the
arsenide compounds show a similar interplay among structure,
antiferromagnetism and superconductivity. These include the
oxyarsenide systems obtained through replacing lanthanum by other
rare-earth elements such as Ce, Pr, Nd, Sm, and
Gd~\cite{GFChen:08,Ren:08,XHChen:08,Cheng:08}, as well as
oxygen-free arsenides, such as BaFe$_{\rm 2}$As$_{\rm
2}$~\cite{Rotter:08} and SrFe$_{\rm 2}$As$_{\rm
2}$~\cite{Krellner2:08}.

The existence of the antiferromagnetic state naturally raises
the possibility of carrier-doping-induced quantum phase transitions 
in the 
iron pnictides~\cite{Si:08,Fang:08,Xu:08},
but the situation is not yet certain. 
Theoretically, the evolution of the Fermi surface as a function 
of carrier doping is not yet well understood, and this 
limits the study of quantum criticality.
Experimentally,
earlier measurements
in LaO$_{\rm1-x}$F$_{\rm x}$FeAs~\cite{Kamihara:JACS08}
and SmO$_{\rm
1-x}$F$_{\rm x}$FeAs~\cite{Liu:08} show a moderate suppression
of the magnetic/structural transition temperature(s) as $x$ is increased;
beyond $x$ of about $\sim 7\%$, the transitions are interrupted by
superconductivity. Further experiments have led to conflicting
reports for the first-order or second-order nature of the
carrier-induced zero-temperature 
magnetic and structural
phase transitions~\cite{Luetkens:08,Drew:08,Zhao:08}.

We propose that an alternative to a possible doping-induced quantum 
phase transition is one that is accessed by changing the relative 
strength of electron-electron correlations. Thus we suggest that 
the iron pnictides may exhibit a new example and setting for quantum
criticality. Our approach is motivated by the phenomenological and
theoretical evidence that the parent iron pnictide is a 
``bad metal''~\cite{Si:08,Haule:08,Laad:08}. 
Accordingly,
we formulate
our considerations in terms of an {\it incipient} Mott insulator:
the electron-electron interactions 
lie close to, but do not exceed the critical value for the insulating state. 
Within this picture, the electronic excitations comprise an
incoherent part away from the Fermi energy, and a coherent part in its
vicinity. The incoherent electronic excitations are described
in terms of localized Fe magnetic moments, with frustrating superexchange
interactions. The latter have been discussed earlier by two of us~\cite{Si:08} 
and others~\cite{Yildirim:08}. This division of the electron spectrum 
is a simple and convenient way of analyzing the complex behavior 
of a bad metal close to the Mott transition, whose spectrum exhibits
incipient upper and lower Hubbard bands and a coherent quasiparticle
peak at the Fermi energy~\cite{Georges:96}.

The coupling of the local moments to the coherent electronic excitations 
competes against the magnetic ordering. A magnetic quantum critical point
arises when the spectral weight of the coherent electronic excitations 
is increased to some threshold value.

\section{The electron spectrum}
\begin{figure}[h]
\center\includegraphics[totalheight=0.23\textheight, viewport=50 100 800 470,clip]
{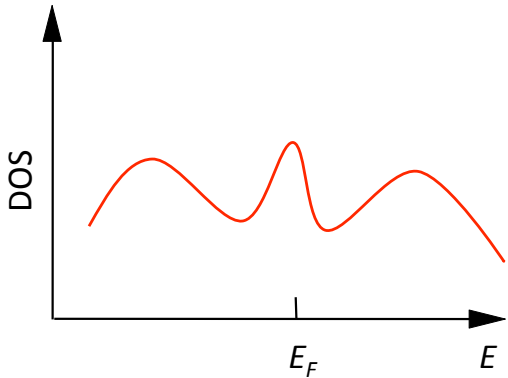}
\caption
{{\bf Single-electron spectral function as the sum of
coherent and incoherent parts.}
The single-electron density
of states (DOS) is plotted against energy ($E$); $E_F$ is the 
Fermi energy.
Each peak may
contain additional structure 
due to the multi-orbital nature of the iron
pnictides. 
The percentage of the total 
spectral weight that belongs to the coherent part 
is defined as $w$,
which goes from $1$ when the interaction is absent,
to $0$ when the interaction reaches and goes beyond the 
Mott-transition threshold.}
\label{dos}
\end{figure}

The incoherent and coherent parts of the single-electron spectral
function are illustrated schematically in Fig.\ \ref{dos}.
The central peak describes the coherent itinerant carriers;
these are the electronic excitations that are responsible for a Drude
optical response and that are adiabatically connected
to their non-interacting counterparts.
The side peaks describe the incoherent excitations, 
vestiges of the lower and upper Hubbard bands associated with 
a Mott insulator that would arise if the electron-electron
interactions were larger than the Mott-localization threshold.
Each of the three peaks may in general have a complex structure due to the 
multi-orbital nature of the iron pnictides.
The decomposition of the electronic spectral weight into 
coherent and incoherent parts is natural for a metal near a Mott 
transition~\cite{Allen:06,Georges:96}.

We use $w$ to denote the percentage of the
spectral weight lying in the coherent part of the spectrum.
A relatively small $w$ may be inferred for the iron pnictides,
because the Drude weight seen in the optical 
conductivity~\cite{Dong:08,Boris:08,Hu:08}
is very small (on the order of 5\% of the total spectral weight integrated to 
about 2~eV).  A small $w$ corresponds to an interaction strength sufficiently
large that the system is close to the Mott transition, albeit on the metallic
side; this implies a large electron-electron scattering rate, consistent with
the observed large electrical resistivity  (on the order of 
$0.5~{\rm m \Omega \cdot cm}$ for single crystals
and $5~{\rm m \Omega \cdot cm}$ for polycrystals) at room temperature.
In terms of electrical conduction, the iron pnictides are similar to,
{\it e.g.}, ${\rm V_2O_3}$, a bad metal (with a room temperature resistivity~\cite{V2O3} 
of about 0.5~${\rm m \Omega \cdot cm}$) that is known to be on the verge of a Mott
transition, and is very different from, {\it e.g.},  ${\rm Cr}$, a simple metal 
(with a room temperature resistivity~\cite{Cr} of about 0.01~${\rm m \Omega \cdot cm}$)
which orders into a spin-density-wave ground state.

\section{Effective Hamiltonian}

To study the magnetism, the incoherent spectrum is naturally described in terms of 
localized magnetic moments, leading to a matrix $J_1$--$J_2$ model~\cite{Si:08}:
\begin{eqnarray}
H_J  &=&
\sum_{\langle ij\rangle} J_1^{\alpha\beta}
{\bf s}_{i,\alpha} \cdot {\bf s}_{j,\beta}
+
\sum_{\langle\langle ij\rangle\rangle} J_2^{\alpha\beta}
{\bf s}_{i,\alpha} \cdot {\bf s}_{j,\beta} \nonumber \\
&+&J_H \sum_{i,\alpha \ne \beta} {\bf s}_{i,\alpha} \cdot {\bf s}_{i,\beta} .
\label{H_J}
\end{eqnarray}
Here, $J_1$
and $J_2$
label the superexchange interactions
between two nearest-neighbor (n.n.,$\langle ij\rangle$) and
next-nearest-neighbor (n.n.n.,$\langle\langle ij\rangle\rangle$)
Fe sites, respectively.
Both are matrices in the orbital
basis, $\alpha, \beta$ with these indices summed when 
repeated. $J_H$ is the Hund's coupling.

Eq.~(\ref{H_J}) reflects the projection of the full interacting problem
to the low-energy subspace when the system is a Mott insulator ($w=0$)
and the single-electron excitations have only the incoherent part.
When the single-electron excitations also contain the coherent
part ($w$ being non-zero but small,
see Fig.~\ref{dos}),
these coherent electronic excitations
must be included in the low-energy subspace as well. 

We will use the projection procedure of Ref.~26 to
construct the effective low-energy Hamiltonian.
We denote by $d_{{\bf k}\alpha\sigma}^{coh}$ the $d$-electron operator projected to the coherent part of the electronic states near the Fermi energy,
and define the incoherent part through $d_{{\bf k}\alpha\sigma} \equiv d_{{\bf k}\alpha\sigma}^{coh} + d_{{\bf k}\alpha\sigma}^{incoh}$.
Therefore, unlike the full $d$-electron operator,
$d_{{\bf k} \alpha \sigma}^{coh}$ does not satisfy the fermion anticommutation rule. Indeed, its spectral function integrated over frequency
defines $w$. We therefore introduce
$c_{{\bf k} \alpha \sigma}=(1/\sqrt{w})d_{{\bf k} \alpha \sigma}^{coh}$,
so that $c_{{\bf k} \alpha \sigma}$
has a total spectral weight of $1$ and
satisfies $\{c_{{\bf k} \alpha \sigma}, c^{\dagger}_{{\bf k}\alpha \sigma}\}=1$.

We then have the effective low-energy Hamiltonian terms for
the coherent itinerant carriers ($H_c$) and for their mixing 
with the local moments ($H_m$):
\begin{eqnarray}
H_c &= &\sum_{{\bf k},\alpha,\sigma} \epsilon_{{\bf k}\alpha\sigma}
c_{{\bf k} \alpha \sigma}^{\dagger}
c_{{\bf k} \alpha \sigma}
= w \sum_{{\bf k},\alpha,\sigma} E_{{\bf k}\alpha\sigma}
c_{{\bf k} \alpha \sigma}^{\dagger}
c_{{\bf k} \alpha \sigma}
\nonumber\\
H_m &=& \sum_{{\bf k} {\bf q} \alpha \beta \gamma} g_{{\bf k},{\bf
q} \alpha \beta \gamma}~ c_{{\bf k}+{\bf q}\alpha\sigma}^{\dagger}
\frac{\boldsymbol\tau_{\sigma\sigma^{\prime}}}{2}
c_{{\bf k}\beta\sigma^{\prime}} \cdot {\bf s}_{\mathbf{q}\gamma} \nonumber\\
&=& w \sum_{{\bf k} {\bf q} \alpha\beta\gamma} G_{{\bf k},{\bf
q}\alpha\beta\gamma}~ c_{{\bf k}+{\bf q}\alpha\sigma}^{\dagger}
\frac{\boldsymbol\tau_{\sigma\sigma^{\prime}}}{2} c_{{\bf
k}\beta\sigma^{\prime}} \cdot {\bf s}_{\mathbf{q}\gamma} \;.
\label{H0_Hcouple}
\end{eqnarray}
Here $\boldsymbol\tau$ labels the three Pauli matrices.
In the projection procedure leading to Eq.\ (\ref{H0_Hcouple}),
we keep $d_{{\bf k}\alpha\sigma}^{coh}$ as part of the low-energy
degrees of freedom; the prefactor $w$ in the first equation comes
from the rescaling $c_{{\bf k} \alpha \sigma}
=(1/\sqrt{w})d_{{\bf k} \alpha \sigma}^{coh}$ and $E_{{\bf k}\alpha\sigma}$ is therefore the conduction-electron dispersion at $w=1$.
At the same time, we integrate out the high energy states involved
in $d_{{\bf k}\alpha\sigma}^{incoh}$. To the leading order in $w$, this procedure is carried out at the $w=0$ point which is
taken to have a full gap~\cite{Moeller:95}; as a result, the effective coupling
$G_{{\bf k}{\bf q}\alpha\beta\gamma}$ is of order $w^0$.
Beyond the leading order in $w$, the coupling constants will acquire further corrections. The computation of these corrections
is difficult, since, at those orders, the spectrum becomes
continuous from the coherent to incoherent part (cf.~Fig.~\ref{dos});
it is left for future work. Still, our leading-order analysis
captures the form of the low-energy effective Hamiltonian,
which is
\begin{equation}
H_{eff} = H_J + H_c + H_m\; .
\label{H-low}
\end{equation}

\section{J$_1$--J$_2$ competition}
\begin{figure}[t]
\includegraphics[totalheight=0.2\textheight, viewport=50 150 800 480,clip]
{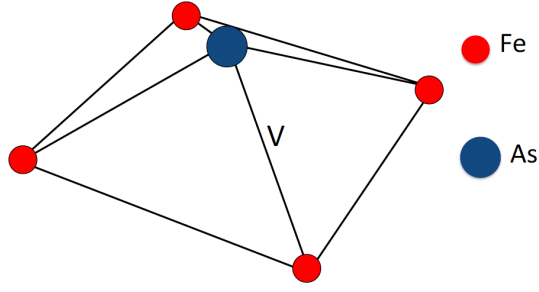}
\caption{\label{Fe-plaquette}
{\bf A square plaquette of Fe ions, with an As ion sitting above
or below the center of the plaquette.}
The hybridization
term is
written as $h_V = \sum_{{\bf r}}
V[p^{\dagger}_{\sigma}({\bf r}) 
\sum_{\square}d_{\sigma}({\bf r})
+ h.c.]$,
and the energy levels for the Fe 3$d$ orbital
and As 4$p$ orbital are 
$\epsilon_d$
and $\epsilon_p$,
respectively.
The resulting superexchange interaction is
$h_J
= [2V^4/(\epsilon_p-\epsilon_d)^3] 
\sum_{\bf r}
[\sum_{\square} {\bf s}({\bf r})]^2$.
} 
\end{figure}

The superexchange interactions in the iron pnictides contain n.n.\ and
n.n.n.\ terms because of the specific relative locations of the ligand As atoms
and Fe atoms~\cite{Si:08,Yildirim:08,Ma:08}. To assess the tunability 
of $J_1$ and $J_2$, we consider an over-simplified case, illustrated 
in Fig.\ \ref{Fe-plaquette}. Here, only one Fe 3$d$ orbital is considered. 
We assume that the 3$d$ orbital on each of the
four corners of a square plaquette has an identical hybridization matrix element,
$V$, with one As 4$p$ orbital located above the center of the plaquette. 
The superexchange interaction is found to be $h_J \propto \sum_{\bf r}
[\sum_{\square} {\bf s}({\bf r})]^2$,
where ${\bf r}$ labels a plaquette in the 2D square lattice and the summation 
$\sum_{\square}$ 
is over the four Fe sites of a plaquette. For classical spins, this is
the canonical case of magnetic frustration: all states with
$\sum_{\square} {\bf s}({\bf r}) = 0$ 
are degenerate. Written in the form of Eq.~(\ref{H_J}), this corresponds 
to $J_2=J_1/2$. This discussion is instructive for the understanding of the 
realistic exchange interactions in the iron pnictides. Several aspects are
neglected in the simplified analysis given above. First, multiple $3d$ orbitals
are important, and the hybridization is orbital-sensitive. Both the $J_1$ and 
$J_2$ interactions are therefore matrices. Second, the real band structures 
must be described by more complex $d$-$p$,  $p$-$p$, and $d$-$d$ tight-binding
parameters. Both features spoil the elementary $J_2=J_1/2$ relationship. 
Still, the simple considerations given above suggest that the overall 
strength of $J_2$ and $J_1$  -- {\it i.e.} the largest eigenvalues
of the $J_2$ and $J_1$ matrices -- are comparable with each other.
Detailed analysis of the matrix elements indicates that there are more 
entries in the $J_2$ matrix than in the $J_1$ matrix that correspond to
the dominating antiferromagnetic component, and that the overall magnitude
(the largest eigenvalue) of the $J_2$ matrix will be somewhat larger than 
half of that of the $J_1$ matrix. This conclusion is supported by the fitting
of the {\it ab initio} results of the ground state energies for various magnetic
configurations in terms of $J_1$ and $J_2$ parameters in the (non-matrix) 
Heisenberg form~\cite{Yildirim:08,Ma:08}. This range of $J_2/J_1$ leads 
to a two-sublattice collinear antiferromagnetic ground state, consistent
with the results of the neutron scattering experiment~\cite{Cruz:08}.

On the one hand, the above argument implies that the magnetic frustration
effect is strong, and can provide significant quantum fluctuations leading
to a reduced ordered moment. On the other hand, it suggests that the degree
to which $J_2/J_1$ can be tuned in practice could be limited.

\section{Magnetic quantum critical point}
The order parameter for the two-sublattice antiferromagnet
appropriate for $J_2/J_1 > 1/2$ is the staggered
magnetization, ${\bf m}$, at wave vector ${\bf Q}=(\pi,0)$. The effective theory
for the $H_J$ term alone corresponds to a $\phi^4$ theory whose action is of the form
${\cal S} \sim r\phi^2 +  u\phi^4$.
The coupling to the coherent quasiparticles is given by given by the $H_m$ term of Eq.\ (\ref{H0_Hcouple}); it
 causes a shift 
of the tuning parameter $r$ and also introduces a damping term.
These contributions to the $r$ coefficient are given by:
\begin{eqnarray}
\Delta r + i \Gamma = 
\sum_{\bf k,\alpha,\beta,\gamma} 
g_{{\bf k}{\bf q}\alpha\beta\gamma}^2 
a_{\gamma}^2
\frac{ f(\epsilon_{{\bf k}+{\bf
q},\alpha}) -f(\epsilon_{{\bf k},\beta})} 
{i\omega_n -(\epsilon_{{\bf k}+{\bf q},\alpha}
- \epsilon_{{\bf k},\beta})}
~. 
\label{damping}
\end{eqnarray}
Here $f(\epsilon)$ is the Fermi-Dirac distribution function
and $a_{\gamma}$ is an orbital-dependent coefficient:
$\sum_{\gamma} a_{\gamma} {\bf s}_{\gamma}$ appears 
in the order parameter for the $(\pi,0)$ antiferromagnet.
Note that both $g_{{\bf k},{\bf q}\alpha\beta\gamma}$
and $\epsilon_{{\bf k},\beta}, \epsilon_{{\bf k}+{\bf q},\alpha}$ 
are linear order in $w$. 
We can infer from Eq.~(\ref{damping})
that the damping term is
of the order $w^0$ at low energies:
for $|\omega_n| \ll w W$ ($W$ is the bandwidth), 
$\Gamma = \gamma |\omega_n| $,
where $\gamma$ is, to leading order in $w$,
the constant value associated with the
couplings
and density of states of the $w=1$ case.
Note that $\gamma$ is non-zero because, for the parent 
compounds, ${\bf Q}$ connects the hole pockets near
the $\Gamma$ point of the Brillouin zone (BZ) and 
the electron pockets near the
M points (in the unfolded BZ notation). At the same time,
$\gamma $ does not diverge since the nesting is not perfect.
The existence of the linear in $\omega$ 
damping term is in contrast
to the doped case, where ${\bf Q}$ no longer connects the hole- and
electron- Fermi surfaces~\cite{Xu:08}.
Importantly, we can also infer from Eq.~(\ref{damping})
that the leading frequency and temperature independent term 
$\Delta r = w A_{\bf Q}$ 
is linear in $w$,
with 
$A_{\bf Q} =
\sum_{{\bf k},\alpha,\beta,\gamma} 
G_{{\bf k}{\bf q}\alpha\beta\gamma}^2 
a_{\gamma}^2 [\Theta(E_F-E_{{\bf k}+{\bf Q}})
- \Theta(E_F-E_{{\bf k}})]/(E_{{\bf k},\beta}
-E_{{\bf k}+{\bf Q},\alpha})$
(where $\Theta$ is the Heaviside function)
is independent of $w$, and positive.

\begin{figure}
\includegraphics[totalheight=0.2\textheight, viewport=40 100 800 430,clip]
{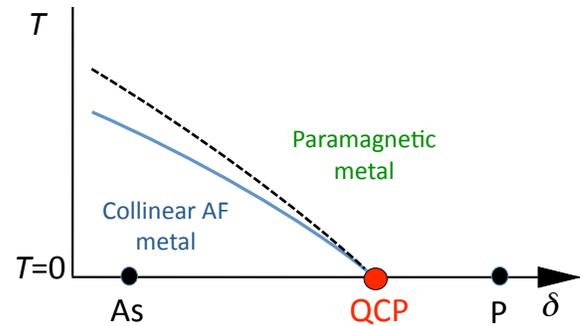}
\caption{\label{qpt}
{\bf Magnetic quantum phase transition in the parent compounds
of the iron pnictides.} The blue solid/black dashed lines represent
the magnetic/structural transitions, respectively.
$\delta$ is a non-thermal control parameter: increasing $\delta$
enhances the spectral weight in the coherent 
part of the single-electron excitations ({\it cf.}
Fig.~\ref{dos}).
The QCP, at $\delta=\delta_c$, separates
a two-sublattice collinear AF ground state from a paramagnetic one.
A specific example for $\delta$ is
the concentration of P doping for As:
a parent iron pnictide with As is an antiferromagnetic metal,
whereas its counterpart with P is non-magnetic;
the possibility that the latter is superconducting is not 
shown in the phase diagram.
}
\end{figure}

The low energy Ginzburg-Landau theory then takes the form,
\begin{eqnarray}
{\cal S}\! & = &\!\! \int d {\bf q} 
\! \! \int \!\! d \omega 
\left[ r(w) + c ({\bf q} -{\bf Q})^2 +\omega^2 + \gamma |\omega| \right]
[\, {\bf m} ({\bf q}, \omega ) \,]^2
\nonumber\\
&+&\!\!\, u \, \prod_{i=1}^{4} \int \!\! d {\bf q}_i \! \! \int \!\! d \omega _i
\, \delta\!\left(\sum_i {\bf q}_i\right) \delta\!\left(\sum_i \omega _i\right)
[\, {\bf m} \,]^4 + \ldots ,
\label{S-effective}
\end{eqnarray}
where $r(w)=r(w=0)+w A_{\bf Q}$.
$r(w=0)$ is negative, placing the system at $w=0$ 
to be antiferromagnetically ordered. The linear
in $w$ shift, $wA_{\bf Q}$, causes $r(w)$ to vanish at a $w=w_c$,
leading to a quantum critical point.
In terms of the external control parameter $\delta$,
shown in Fig.\ \ref{qpt}, 
$w=w_c$ defines $\delta=\delta_c$.
The $\phi^4$ theory describes a $z=2$ (where $z$ is the dynamical
exponent) antiferromagnetic
quantum phase transition, which is generically second order.

The $O(3)$ vector ${\bf m}$, corresponding to the $(\pi,0)$ order, is
accompanied by another $O(3)$ vector, ${\bf m}'$ that describes the
$(0,\pi)$ order. These two vector order parameters accommodate a
composite scalar, ${\bf m} \cdot {\bf m}'$, the order
parameter for an Ising transition~\cite{Chandra,Fang:08,Xu:08}.
In turn, the Ginzburg-Landau action,
Eq.~(\ref{S-effective}), contains 
a quartic 
coupling $\tilde{u}({\bf m} \cdot {\bf m}')^2$
[as well as $u'{\bf m}^2({\bf m'})^2$]. In the $z=2$ case
here, the $\phi^4$ theory is at effective dimension
$d+z=4$. At the QCP of the $O(3)$ transition, the $\tilde{u}$ quartic coupling
term is marginally relevant
in the renormalization group sense.
The $T=0$ transition could therefore either be turned to first 
order, or be split into two continuous transitions,
one for the Ising transition, whose scalar order parameter is  $\vec{m}\cdot\vec{m'}$, (which corresponds to the structural distortion when it is coupled to some structural degrees of freedom) the other for the $O(3)$ magnetic
one. Either effect is expected to be weak, due to the marginal nature
of the coupling. 

The magnetic quantum criticality will strongly contribute to
the electronic and magnetic properties
in the quantum critical regime. We note that since $d=z=2$, there are (marginal) logarithmic corrections to simple Gaussian critical behavior~\cite{hilbert}. 
Following discussions in, {\it e.g.}, Ref.\ \cite{hilbert}, we expect that the specific-heat coefficient will be $C/T \sim \ln (1/T)$,
the NMR relaxation rate $1/T_1 \propto {\rm const}$, and (in the presence of disorder scattering
that smears the Fermi surface) the resistivity
$\rho\propto T$.

\section{Tuning parameter and variation of magnetic order}
The parent materials of the different iron arsenides
will have different internal pressures and ``$c/a$" ratios,
and will correspondingly
have different ratios of the electron-electron interaction
to the effective bandwidth. 
According to our theory, the resulting variation of the coherent spectral weight $w$
will, in turn,
tune the control parameter $r$ in Eq.~(\ref{S-effective}),
and the ordered moment will change accordingly across the different compounds.

Neutron scattering experiments have indeed found 
that the ordered moment
does vary across the parent arsenides. The moment associated
with
Fe-ordering at low temperatures is about 0.2-0.3$\mu_B$/Fe 
in NdOFeAs~\cite{Qiu:08,Chen_Y:08},
0.4$\mu_B$/Fe in LaOFeAs~\cite{Cruz:08},
0.5$\mu_B$/Fe in PrOFeAs~\cite{Kimber:08,Zhao_PrFeAsO},
and 0.8-1.0 $\mu_B$/Fe in CeOFeAs~\cite{Zhao:08},
BaFe$_2$As$_2$~\cite{Huang:08},
and SrFe$_2$As$_2$~\cite{Jesche:08}.

\section*{As$_{1-\delta}$P$_{\delta}$ Series of the Parent Iron Pnictides} 
Since the $c$-lattice constant in LaOFeP is smaller than that in LaOFeAs, 
these considerations suggest that the coherent-electron spectral weight 
of the iron phosphides is larger than that of the iron arsenides. 
A consequence is that, in contrast to the arsenide, the phosphide does
{\it not} have a magnetic transition.~\cite{Kamihara:08prb} We then propose that 
a parent iron pnictide series created by P doping of As
presents a means to unmask a magnetic 
quantum critical point.
Our purpose is better served the weaker the superconductivity
is in the P end material. LaOFeAs$_{1-\delta}$P$_{\delta}$ is
promising,
since LaOFeP is a weak superconductor
whose $T_c$ is only a few Kelvin or may even vanish~\cite{Kamihara:06,Maple,Cava}.
CeOFeAs$_{1-\delta}$P$_{\delta}$ may also be of interest in this
context.
While CeOFeAs~\cite{Zhao:08} is antiferromagnetic,
CeOFeP is a paramagnetic metal~\cite{Bruning:08}.
We remark in passing that P-doping for As is more
advantageous than external pressure, because the latter is known to cause a volume collapse \cite{canfield}. It would be interesting to search for a substitution for As such that $w$ could be reduced, leading towards to the Mott insulating state.

To understand the tuning of the microscopic
electronic parameters, we have carried out 
density-functional-theory (DFT) calculations 
on
both CeOFeAs and CeOFeP for comparison. We find that
the $d$-$p$ hybridization matrix is larger in
CeOFeP than
in CeOFeAs. This is consistent with the qualitative consideration
that, compared to
CeOFeAs,
CeOFeP has a higher internal pressure and, hence, 
a higher kinetic energy and smaller ratio of the 
interaction to the bandwidth, thus a larger coherent weight $w$.

\section{Comparison with DFT studies}
We have considered the mechanism for quantum fluctuations
having in mind the proximity to the Mott limit, where the
{\it instantaneous} atomic moment is large (a few $\mu_B$/Fe)
to begin with. Most DFT calculations have shown that the
{\it ordered} moment in the antiferromagnetic ground state
is large, of the order $2\mu_B$/Fe. Moreover, such a large
ordered moment was found within DFT not only for the parent iron pnictides,
but also for their doped counterparts.

Since DFT calculations neglect quantum fluctuations, we are
tempted to interpret the large DFT-calculated moment as essentially
the instantaneous atomic moment. Quantum fluctuations will then
lead to a reduced ordered moment in the true ground state.
The $J_1$--$J_2$ competition together with the coupling of the local
moments to the
coherent itinerant electronic excitations arising naturally
in the Mott-proximity picture we have described, is just such
a mechanism for quantum fluctuations. 
\section{Discussion}
We have developed a framework to describe the quantum magnetism 
of the iron pnictides, appropriate for electron-electron interactions
that are of an intermediate strength to place the materials at the 
delicate boundary between itinerancy and localization. 
Our description takes into account the interplay between the
itinerant and local-moment aspects, which are naturally associated with
the interaction-induced coherent and incoherent parts of the 
electronic excitations. Enhancement of the spectral weight associated
with the coherent electronic excitations weakens the magnetic order,
and induces a magnetic quantum critical point. Our characterization
of the magnetic excitations is important not only for the understanding
of the existing and future experiments in the normal state,
but also for the microscopic understanding of high temperature
superconductivity in the iron pnictides and related metallic systems
close to a Mott transition. In addition, realization of a magnetic
quantum critical point in the iron pnictides provides
a new setting to explore some of the rich 
complexities~\cite{Sachdev_natphys,Gegenwart_natphys} of
quantum criticality; this is much-needed since quantum critical points have so far been explicitly observed only in a very
small number of materials.

\begin{acknowledgments}
We thank
G. Cao, P. Coleman, C. Geibel, A. Jesche, C. Krellner, Z.-Y. Lu,
E. Morosan, D. Natelson, C. Xu, and Z.A. Xu for useful discussions.
This work has been supported by 
the NSF of China and PCSIRT
of Education
Ministry of China (J.D.), the NSF
Robert A. Welch Foundation (Q.S.), and the U.S. Department of Energy
(J.-X.Z.).
\end{acknowledgments}

\end{article}

\end{document}